\documentclass[aps,preprint]{revtex4}%
\usepackage{amsfonts}
\usepackage{amsmath}
\usepackage{amssymb}
\usepackage{graphicx}%
\providecommand{\U}[1]{\protect\rule{.1in}{.1in}}

\newcommand{\beq}{\begin{equation}}
\newcommand{\eeq}{\end{equation}}
\newcommand{\bea}{\begin{eqnarray}}
\newcommand{\eea}{\end{eqnarray}}
\begin{document}
\preprint{ }
\title{Finite Temperature Scaling, Bounds, and Inequalities for the Non-interacting
Density Functionals}
\author{James W. Dufty}
\affiliation{Department of Physics, University of Florida, Gainesville FL 32611}
\author{S.B. Trickey}
\affiliation{Quantum Theory Project, Dept.\ of Physics and Dept.\ of Chemistry, University
of Florida, Gainesville FL 32611}

\begin{abstract}
Finite temperature density functional theory requires representations for the
internal energy, entropy, and free energy as functionals of the local density
field. A central formal difficulty for an orbital-free representation is
construction of the corresponding functionals for non-interacting particles in
an arbitrary external potential. That problem is posed here in the context of
the equilibrium statistical mechanics of an inhomogeneous system. The density
functionals are defined and shown to be equal to the extremal state for a
functional of the reduced one-particle statistical operators. Convexity of the
latter functionals implies a class of general inequalities. First, it is shown
that the familiar von Weizs\"acker lower bound for zero temperature
functionals applies at finite temperature as well. An upper bound is obtained
in \ terms of a single-particle statistical operator corresponding to the
Thomas-Fermi approximation. Next, the behavior of the density functionals
under coordinate scaling is obtained. The inequalities are exploited to obtain
\ a class of upper and lower bounds at constant temperature, and a
complementary class at constant density. The utility of such constraints and
their relationship to corresponding results at zero temperature are discussed.

\end{abstract}
\date{May 11, 2011}
\maketitle

\section{Introduction}

\label{sec1}Density functional theory (DFT) is an important structure for
posing problems of equilibrium many-body physics in a non-perturbative
context. Most applications, with many significant successes, have been in the
extremes of zero-temperature ground state systems
\cite{Yang-Parr-book,DreizlerGrossBook,KryachkoLudena,Eschrig,LudenaKarasiev2002,FilohaisEtAl03,
EngelDreizler11} or high-temperature classical systems \cite{Evans92,
Lowen94,Lutsko10}. Shortly after the ground-state Hohenberg-Kohn-Sham
formulation, a finite-temperature generalization was given \cite{Mermin65} but
it has attracted far less development than the ground-state version
\cite{Dharma95}. Recently, attention has been directed toward systems in a
difficult intermediate domain known as \textquotedblleft warm dense
matter\textquotedblright. Extensions of existing density functional approaches
(including procedures to develop approximate functionals non-empirically) to
this domain have met with challenges. One is the temperature-dependence itself
of the functionals. Another is technical, associated with solving a
self-consistent single-particle eigenvalue problem (the Kohn-Sham equation)
for which a growing number of eigenvalues and eigenfunctions (orbitals) is
required as the temperature is increased at solid densities. For many
applications, such as determining forces for Born-Oppenheimer molecular
dynamics of ions in warm, dense matter, this growth produces a computational
bottleneck. The Kohn-Sham procedure is not intrinsic to DFT, so an alternative
approach is to return to the expression of the entire free energy as a
functional of the density. Once known, the extremum condition of DFT becomes a
single non-linear, non-local equation for the density whose solution replaces
that constructed from the Kohn-Sham orbitals and eigenfunctions. Solving that
Euler equation is facilitated by introducing the Kohn-Sham separation into a
non-interacting functional, which is presumably dominant, and the remainder,
but without introducing orbital-dependent expressions. This approach is known
as orbital free DFT (OFDFT). An example is given by the Thomas-Fermi-Dirac
model, which is a local density approximation.

A critical component for OFDFT is the free energy density functional for an
inhomogeneous but non-interacting system. The objective here is to define the
free energy, energy, and entropy density functionals for a non--interacting
system at finite temperature in forms suitable for constructing
approximations. (Note the correspondence with the ground state theory, to wit,
the definition of the Kohn-Sham kinetic energy and exact exchange energy.) To
aid in that construction important exact constraints are obtained here as
well. These include scaling laws, upper and lower bounds, and inequalities.
Development and use of such constraints is promising in light of the history
of analogous developments in zero-temperature DFT.

The approach here is to exploit the relationship of DFT to the equilibrium
statistical mechanics of non-uniform systems \cite{Thermo}. The variational
context of DFT is simply related to the extremal properties of the ensembles
representing equilibrium systems. In the next section, thermodynamics is
described in the grand ensemble, for which the thermodynamic variables are the
temperature $T$ and chemical potential $\mu$. The presence of an external
single-particle potential $v(\mathbf{r})$ transforms the chemical potential to
a local function $\mu\left(  \mathbf{r}\right)  =\mu-v(\mathbf{r})$. A change
of thermodynamic variables $T,\mu\left(  \mathbf{r}\right)  $ to temperature
and density $T,n\left(  \mathbf{r}\right)  $ is effected by an appropriate
Legendre transformation. The resulting thermodynamic functionals of
$T,n\left(  \mathbf{r}\right)  $ are the desired density functionals for DFT.
The extremum condition of DFT is simply the thermodynamic identity relating
the functional derivative of the free energy to the chemical potential.

This context for DFT is described briefly in the next section and then
specialized to the non-interacting system in section \ref{sec3}. The equation
for the density provides the definition of the local chemical potential as a
functional of the density. The free energy, energy, and entropy are
represented in terms of the equilibrium single-particle reduced density
matrix, or Fermi operator. The latter depends on the chemical potential, so
that in this way these properties are identified as functionals of the density
through the chemical potential. Next, corresponding functionals of an
arbitrary single-particle statistical operator are defined. Using convexity of
the free energy functional of statistical operators, it is shown that the
minimum occurs for the Fermi operator, resulting in the free energy density
functional being a minimum for the free energy statistical operator
functionals. This is the source for the inequalities described subsequently.

In section \ref{sec4}, the transformation of the density functionals under
coordinate scaling is obtained from dimensional analysis. An alternative
constructive derivation which connects explicitly with ground-state uniform
scaling \cite{LevyPerdew85,Levy87,Levy91} is given in Appendix \ref{apB}.
Next, the familiar von Weizs\"acker lower bound on the kinetic energy for
zero-temperature DFT \cite{SearsParrDinur80,Harriman85,Herring86} is shown to
hold for finite temperature as well. A statistical operator underlying the
Thomas-Fermi approximation for the density is defined via an appropriate
Wigner function representation and used to obtain an upper bound for the free
energy. Finally, two classes of inequalities are established for all density
functionals using the scaling laws. One is for fixed temperature at different
densities, while the other is for fixed density and different temperatures.
For the case of differential scaling, the results reduce to differentials of
equilibrium thermodynamics.

In the last section these results are discussed and their utility for
constraint-based approximations described.

\section{Equilibrium statistical mechanics for a non-uniform system}

\label{sec2} While treatment of a homogeneous non-interacting Fermion system
is a textbook staple, the statistical mechanics of \textit{in}-homogeneous
non-interacting systems is less well documented. The formal structure of
ensemble DFT has been discussed in diverse forms
\cite{GuptaRajagopal82,PPLB82,Perdew85,StoitsovPetkov88,Dreizler89,Eschrig10}
since Ref.\ \onlinecite{Mermin65}. Some of those formulations resemble the
ground-state theory more than others. Homogeneous scaling at finite
temperature has been treated recently in a manner closely related to
ground-state scaling \cite{PittalisEtAl11}. The present approach is
complementary, in the sense of providing a close, systematic connection with
statistical mechanical methods widely used in the treatment of electronically
hot systems, \textit{e.g.} plasmas. To set notation and make the context
explicit, we begin with a brief summary of the relevant statistical mechanics.

The system of interest is comprised of $N$ electrons in a uniform, rigid,
neutralizing background (\textquotedblleft one component
plasma\textquotedblright, \textquotedblleft jellium\textquotedblright,
\textquotedblleft electron gas\textquotedblright) \cite{Giuliani05}, and in an
external potential $v$. The Hamiltonian is
\begin{equation}
\widehat{H}=\widehat{H}_{e}+\sum_{\alpha=1}^{N}v(\widehat{\mathbf{q}}_{\alpha
}). \label{2.1}%
\end{equation}
A caret over a symbol indicates that it is an operator. Here
\begin{equation}
\widehat{H}_{e}=\sum_{\alpha=1}^{N}\frac{\widehat{p}_{\alpha}^{2}}{2m_{\alpha
}}+\tfrac{1}{2}\sum_{\alpha\neq\gamma=1}^{N}\frac{e^{2}}{|\hat{\mathbf{q}%
}_{\alpha}-\hat{\mathbf{q}}_{\gamma}|}-\sum_{\alpha=1}^{N}\int d\mathbf{r}%
\frac{n_{b}e^{2}}{|\hat{\mathbf{q}}_{\alpha}-{\mathbf{r}}|}+\tfrac{1}{2}\int
d\mathbf{r}d\mathbf{r}^{\prime}\frac{n_{b}e^{2}}{|\mathbf{r}-\mathbf{r}%
^{\prime}|}\,. \label{2.2}%
\end{equation}
The integrations extend over the volume $V$ and $n_{b}$ is the average number
density of the background chosen for over-all charge neutrality. The first two
terms are the kinetic energy and electron-electron Coulomb interaction
energies. The third term is the interaction energy of the electrons with the
neutralizing background (charge density $n_{b}e$), while the last term is the
interaction energy among elements of that uniform background. The origin and
necessity for this neutralizing background arises from the long range nature
of the Coulomb potential - a macroscopic equilibrium state for the electrons
alone does not exist \cite{Lieb76}. Charge neutrality is a necessity for
thermodynamic stability, {\it e.g.} systems of both electrons and positive 
ions. The
system considered here is a model for electrons in such two-component systems,
where the uniform, rigid, neutralizing background assures that thermodynamics
exists \cite{Lieb75,Martin88}. The electron density operator is defined as
\begin{equation}
\widehat{n}(\mathbf{r})=\sum_{\alpha=1}^{N}\delta\left(  \mathbf{r}%
-\widehat{\mathbf{q}}_{\alpha}\right)  . \label{2.3}%
\end{equation}
Note that
\begin{equation}
\widehat{n}(\mathbf{r})\widehat{n}(\mathbf{r}^{\prime})=\sum_{\alpha\neq
\gamma=1}^{N}\delta\left(  \mathbf{r}-\widehat{\mathbf{q}}_{\alpha}\right)
\left(  \mathbf{r}^{\prime}-\widehat{\mathbf{q}}_{\gamma}\right)
+\delta(\mathbf{r}-\mathbf{r}^{\prime})\widehat{n}(\mathbf{r}) \label{2.4}%
\end{equation}
Then Eq.\ (\ref{2.1}) becomes
\begin{align}
\widehat{H}  &  =\sum_{\alpha=1}^{N}\frac{\widehat{p}_{i\alpha}^{2}%
}{2m_{\alpha}}\,+\int d\mathbf{r}v(\mathbf{r})\widehat{n}(\mathbf{r}%
)\nonumber\\
&  +\tfrac{1}{2}\int d\mathbf{r}d\mathbf{r}^{\prime}\frac{e^{2}}%
{|\mathbf{r}-\mathbf{r}^{\prime}|}\left[  \left(  \widehat{n}(\mathbf{r}%
)-n_{b}\right)  \left(  \widehat{n}(\mathbf{r}^{\prime})-n_{b}\right)
-\delta({\mathbf{r}}-{\mathbf{r}}^{\prime})n_{b}\right]  \label{2.5}%
\end{align}
The problem of interest is to determine the equilibrium electron density
$n(\mathbf{r})$ and the thermodynamics for this inhomogeneous electron gas.

The thermodynamic properties for this class of inhomogeneous systems are
determined in the grand canonical ensemble as functions of the inverse
temperature $\beta=1/k_{B}T$ and functionals of the local chemical potential
\begin{equation}
\mu(\mathbf{r})=\mu-v(\mathbf{r}). \label{2.6}%
\end{equation}
which combines the external potential $v(\mathbf{r})$ with the constraint on
average total particle number, the constant chemical potential $\mu$. The
fundamental thermodynamic potential for this ensemble, $\Omega(\beta\mid\mu)$,
is related simply to the pressure
\begin{equation}
\Omega(\beta\mid\mu)=-p(\beta\mid\mu)V=-\beta^{-1}\ln\sum_{N=0}^{\infty
}Tr^{(N)}e^{-\beta\left(  \widehat{H} - \int d\mathbf{r}\mu(\mathbf{r}%
)\widehat{n}(\mathbf{r})\right)  }\,. \label{2.7}%
\end{equation}
The trace extends over the Hilbert space for $N$ electrons. The notation
$(\beta\mid\mu)$ indicates variable dependence and functional dependence,
respectively. The thermodynamical average electron density as a function of
$\beta$ and functional of $\mu(r)$ follows directly as
\begin{equation}
n(\mathbf{r,}\beta\mid\mu)=-\frac{\delta\Omega(\beta\mid\mu)}{\delta\mu\left(
\mathbf{r}\right)  }\Big\vert_{\beta}\,. \label{2.8}%
\end{equation}
Similarly, the internal energy is determined from the derivative with respect
to $\beta$
\begin{equation}
U(\mathbf{r,}\beta\mid\mu)=\frac{\partial\beta\Omega(\beta\mid\mu)}%
{\partial\beta}\Big\vert_{\beta\mu\left(  \mathbf{r}\right)  } \label{2.8a}%
\end{equation}

Now consider a change of variables from $\beta,\mu({\mathbf{r}})$ to
$\beta,n({\mathbf{r}})$ by a Legendre transformation to the free energy
\begin{equation}
F(\beta\mid n)=\Omega\left(  \beta\mid\mu\right)  +\int d\mathbf{r}\mu\left(
\mathbf{r}\right)  n(\mathbf{r}) \, , \label{2.9}%
\end{equation}
By construction $F(\beta\mid n)$ is independent of $\mu(r)$ (\textit{i.e.},
$\delta F(\beta\mid n)/\delta\mu(r)=0$). Therefore it is ``universal'' in the
sense that the functional $F(\beta\mid\cdot)$ is the same for all external
potentials, with its argument evaluated at the corresponding different density
fields. In all of the following, only density fields $n(\mathbf{r,}\beta
\mid\mu)$ that can be represented by a chemical potential $\mu(\mathbf{r})$ as
in (\ref{2.7}) and (\ref{2.8}) are considered. These are the $\mu
$-representable physical densities of thermodynamics. The corresponding
ground-state theory requirement on the density is $v$-representability. With
this restriction on the class of densities, Eqs.\ (\ref{2.7}) and (\ref{2.8})
provide the constructive definition for the free energy functional: calculate
$\Omega(\beta\mid\mu)$ as a functional of $\mu$ from its explicit definition;
invert Eq.\ (\ref{2.8}) for the chemical potential as a functional of the
density $\mu(r,\beta\mid n)$; substitute $\mu(r,\beta\mid n)$ into the right
side of (\ref{2.9}).

The functional derivative of $F(\beta\mid n)$ gives the inverse of
(\ref{2.8})
\begin{equation}
\mu(\mathbf{r,}\beta\mid n)=\frac{\delta F(\beta\mid n)}{\delta n\left(
\mathbf{r}\right)  }\Big\vert_{\beta} \label{2.11}%
\end{equation}
This is an identity which expresses the one - one relationship of a
$\mu\left(  \mathbf{r}\right)  $ with a unique density $n\left(
\mathbf{r}\right)  $. However, it also can be viewed as an equation to find
the density associated with a \emph{given} $\mu\left(  \mathbf{r}\right)  $.
For example, consider the case where $\mu_{0}\left(  \mathbf{r}\right)  $ is
given, and the objective is to find the corresponding $n_{0}\left(
\mathbf{r}\right)  .$ This can be accomplished by solving the equation
\begin{equation}
\mu_{0}(\mathbf{r})=\frac{\delta F(\beta\mid n)}{\delta n\left(
\mathbf{r}\right)  }\Big\vert_{\beta}. \label{2.12}%
\end{equation}
The unique solution is $n\left(  \mathbf{r}\right)  =n_{0}\left(
\mathbf{r}\right)  $. This is no longer an identity, but an equation to be
solved: equality holds \emph{only} for the specific density $n_{0}\left(
\mathbf{r}\right)  $.

Up to here the development is simply equilibrium quantum statistical
mechanics. But the same result follows from DFT in an equivalent way. In terms
of the stipulated local chemical potential $\mu_{0}({\mathbf{r}})$, ensemble
constrained search \cite{ConstrainedSearch} enables us to define the
functional
\begin{equation}
\Omega_{\mu_{0}}\left(  \beta\mid n\right)  \equiv F(\beta\mid n)-\int
d\mathbf{r}\mu_{0}\left(  \mathbf{r}\right)  n(\mathbf{r}). \label{2.13}%
\end{equation}
The condition for an extremum of $\Omega_{\mu_{0}}\left(  \beta\mid n\right)
$ gives (\ref{2.12}) as its Euler equation
\begin{equation}
\frac{\delta\Omega_{\mu_{0}}(\beta\mid n)}{\delta n\left(  \mathbf{r}\right)
}\Big\vert_{\beta}=0\text{ \ }\Rightarrow\text{ \ }\mu_{0}(\mathbf{r}%
)=\frac{\delta F(\beta\mid n)}{\delta n\left(  \mathbf{r}\right)
}\Big\vert_{\beta}\,. \label{2.14}%
\end{equation}
From Eqs.\ (\ref{2.9}), (\ref{2.12}) note that
\begin{equation}
\Omega_{\mu_{0}}(\beta\mid n)=\Omega\left(  \beta\mid\mu\right)  -\int
d\mathbf{r}\left(  \mu_{0}\left(  \mathbf{r}\right)  -\mu(\mathbf{r,}\beta\mid
n)\right)  n(\mathbf{r}), \label{2.15}%
\end{equation}
and so%
\begin{equation}
\Omega_{\mu_{0}}(\beta\mid n_{0})=\Omega\left(  \beta\mid\mu_{0}\right)  .
\label{2.16}%
\end{equation}

The case of primary interest is an external potential due to a given
configuration of ions
\begin{equation}
v_{0}(\mathbf{r})=\mu-\mu_{0}(\mathbf{r})=Ze^{2}\int d\mathbf{r}%
d\mathbf{r}^{\prime}\frac{\left(  \widehat{n}(\mathbf{r})-n_{e}\right)
\widehat{n}_{i}(\mathbf{r})}{\left\vert \mathbf{r}-\mathbf{r}^{\prime
}\right\vert },\hspace{0.25in}\widehat{n}_{i}(\mathbf{r})=\sum_{\alpha
=1}^{N_{i}}\delta\left(  \mathbf{r}-\widehat{\mathbf{R}}_{i\alpha}\right)
\label{2.16a}%
\end{equation}
where $Z$ is the ion charge number, and $ZN_{i}=N$. For simplicity, we
discuss only a single ionic species; generalization is obvious.

Determination of the electron density field for given ion configurations
(hence, given $\mu_{0}(\mathbf{r})$) involves constructing $F(\beta\mid n)$
and solving (\ref{2.14}) for $n_{0}({\mathbf{r}})$. The thermodynamic
properties then are determined from $F(\beta\mid n_{0})$. The problem of
constructing $F(\beta\mid n)$ is separated into two parts: finding the
corresponding functional for non-interacting electrons (but in the external
fields of the ions), $F^{(0)}(\beta\mid n)$, and finding the functional
associated with the contribution from Coulomb interactions, $\Delta
F(\beta\mid n)$. This is a logical separation, since $F^{(0)}(\beta\mid n)$ is
itself a universal functional, with properties similar to $F(\beta\mid n)$.
For example, there is a unique functional $\mu^{(0)}(\mathbf{r,}\beta\mid n)$
associated with $F^{(0)}(\beta\mid n)$ for each $n$ given by
\begin{equation}
\mu^{(0)}(\mathbf{r,}\beta\mid n)=\frac{\delta F^{(0)}(\beta\mid n)}{\delta
n\left(  \mathbf{r}\right)  }\Big\vert_{\beta} \, . \label{2.17}%
\end{equation}
The notation $\mu^{(0)}$ specifies that the functional is that for the
non-interacting relationship, with the same chemical potential and external
potential
\begin{equation}
\mu^{(0)}(\mathbf{r,}\beta\mid n)=\mu-v(\mathbf{r}) \, . \label{2.17a}%
\end{equation}

The Coulomb contributions $\Delta F(\beta\mid n)$ (``excess free energy'' in
statistical mechanical usage) conventionally are divided into a mean-field
(Hartree) contribution plus a remainder, the exchange-correlation
contribution
\begin{equation}
\Delta F(\beta\mid n)=\frac{1}{2}\int d\mathbf{r}d\mathbf{r}^{\prime}n\left(
\mathbf{r}\right)  \frac{1}{\left\vert \mathbf{r}-\mathbf{r}^{\prime
}\right\vert }n\left(  \mathbf{r}^{\prime}\right)  +F_{xc}(\beta\mid n),
\label{2.18}%
\end{equation}
so that the total free energy is composed as
\begin{equation}
F(\beta\mid n)=F^{(0)}(\beta\mid n)+\frac{1}{2}\int d\mathbf{r}d\mathbf{r}%
^{\prime}n\left(  \mathbf{r}\right)  \frac{1}{\left\vert \mathbf{r}%
-\mathbf{r}^{\prime}\right\vert }n\left(  \mathbf{r}^{\prime}\right)
+F_{xc}(\beta\mid n). \label{2.19}%
\end{equation}

With these definitions (\ref{2.14}) becomes (with an overall factor of $\beta$
included to make the equations dimensionless)%
\begin{equation}
\beta\mu^{(0)}(\mathbf{r,}\beta\mid n)=\beta\mu_{0}(\mathbf{r})-\beta
e^{2}\int d\mathbf{r}^{\prime}n\left(  \mathbf{r}^{\prime}\right)  \frac
{1}{\left\vert \mathbf{r}-\mathbf{r}^{\prime}\right\vert }-\beta\frac{\delta
F_{ex}(\beta\mid n)}{\delta n\left(  \mathbf{r}\right)  }\Big\vert_{\beta}.
\label{2.20}%
\end{equation}
Again, this relationship does not hold for all $n$. Rather, it is an equation
to be solved for $n=n_{0}$. Its solution can be addressed in two ways. The
Kohn-Sham approach is suggested by combining Eqs.\ (\ref{2.14}), (\ref{2.17}),
and (\ref{2.19}) in a form equivalent to (\ref{2.20}):
\begin{equation}
\frac{\delta F^{(0)}(\beta\mid n)}{\delta n\left(  \mathbf{r}\right)
}\Big\vert_{\beta}\, =\, \mu-v_{KS}(\mathbf{r,}\beta\mid n) \, , \label{2.21}%
\end{equation}
where the KS potential $v_{KS}$ is identified as%
\begin{equation}
v_{KS}(\mathbf{r,}\beta\mid n)=v_{0}(\mathbf{r})+e^{2}\int d\mathbf{r}%
^{\prime}n\left(  \mathbf{r}^{\prime}\right)  \frac{1}{\left\vert
\mathbf{r}-\mathbf{r}^{\prime}\right\vert }+\frac{\delta F_{xc}(\beta\mid
n)}{\delta n\left(  \mathbf{r}\right)  }\Big\vert_{\beta} \, . \label{2.22}%
\end{equation}
Equation (\ref{2.21}) has the form of a non-interacting Fermi system in the
external potential $v_{KS}$. The density and all other thermodynamic
properties then are determined from the self-consistent single-particle
Hamiltonian eigenvalue problem
\begin{equation}
\left(  \frac{\widehat{p}^{2}}{2m}+v_{KS}(\mathbf{r,}\beta\mid n)\right)
\phi_{\alpha}=\epsilon_{\alpha}\phi_{\alpha},\hspace{0.2in}n=\sum_{\alpha
}\left(  e^{\beta\left(  \epsilon_{\alpha}-\mu\right)  }+1\right)
^{-1}\left\vert \phi_{\alpha}\left(  \mathbf{r}\right)  \right\vert ^{2} \, .
\label{2.23}%
\end{equation}
Note that, in this approach, explicit construction of the functionals
$F^{(0)}(\beta\mid n)$ and $\mu^{(0)}(\mathbf{r,}\beta\mid n)$ is avoided, at
the cost of introducing the KS orbital manifold, an issue already mentioned.
The alternative is to construct these functionals, then to solve (\ref{2.20})
directly as an equation for $n$, and use its solution in (\ref{2.19}) to
determine the free energy. This second method is OFDFT; it avoids construction
of the Kohn-Sham eigenfunctions and eigenvalues. Of course, when properly
implemented the two approaches are equivalent.

In the context of OFDFT, the remainder of this discussion is restricted to
determination of $F^{(0)}(\beta\mid n)$ and $\beta\mu^{(0)}(\mathbf{r,}%
\beta\mid n)$. A suitable determination of the corresponding
exchange-correlation functionals is a separate problem not considered here.

\section{Functionals for the non-interacting system}

\label{sec3}

\subsection{Functionals of the density}

Without Coulomb interactions, the Hamiltonian of Eq.\ (\ref{2.4}) is a sum of
single-particle Hamiltonians. Consequently, the evaluation of the grand
potential (\ref{2.7}) can be reduced to a single-particle problem. For
completeness, the analysis is sketched in Appendix \ref{apA}, with the result
\begin{equation}
\Omega^{(0)}(\beta\mid\mu)=-\beta^{-1}Tr\ln\left(  1+e^{-\beta(\frac
{\widehat{p}^{2}}{2m}-\mu(\widehat{\mathbf{q}}))}\right)  ,\hspace{0.2in}%
\mu(\hat{\mathbf{q}})=\mu-v(\hat{\mathbf{q}})\,. \label{3.1}%
\end{equation}
The trace is now taken over the single-particle Hilbert space (spin is not
considered here for simplicity). The density is obtained by (\ref{2.8}) (see
also Appendix \ref{apA})
\begin{equation}
n(\mathbf{r,}\beta\mid\mu)=-\frac{\delta\Omega^{(0)}(\beta\mid\mu)}{\delta
\mu\left(  \mathbf{r}\right)  }\Big\vert_{\beta}=Tr\delta\left(
\mathbf{r}-\widehat{\mathbf{q}}\right)  \widehat{\rho}_{e}, \label{3.2}%
\end{equation}
where the one-body reduced equilibrium statistical operator (hereafter
referred to as the \textquotedblleft Fermi operator\textquotedblright) is
\begin{equation}
\widehat{\rho}_{e}=\left(  e^{-\beta(\frac{\widehat{p}^{2}}{2m}-\mu
(\widehat{\mathbf{q}}))}+1\right)  ^{-1}\,. \label{3.3}%
\end{equation}
The central observation at this point is that (\ref{3.2}) defines the
one-to-one relationship of $\mu(\mathbf{q})$ to an associated density%
\begin{equation}
\mu(\mathbf{q})=\mu^{(0)}(\mathbf{q},\beta\mid n). \label{3.4}%
\end{equation}
A special case is the relationship for a uniform external potential for which
$\mu^{(0)}(\mathbf{q},\beta\mid n)\rightarrow\mu_{e}^{(0)}(\beta,n)$, which is
the familiar relationship of chemical \ potential to density for the ideal
Fermi gas. The functional $\mu^{(0)}(\mathbf{q},\beta\mid\cdot)$ is universal,
\textit{i.e.} it is the same for all external potentials. Use of (\ref{3.4})
in (\ref{3.3}) expresses the Fermi operator as a universal (operator-valued)
functional of the density as well
\begin{equation}
\widehat{\rho}_{e}=\widehat{\rho}_{e}(\beta\mid n)=\left(  e^{\beta
(\frac{\widehat{p}^{2}}{2m}-\mu^{(0)}(\widehat{\mathbf{q}},\beta\mid
n))}+1\right)  ^{-1}\,. \label{3.5}%
\end{equation}

The free energy for the non-interacting system is obtained from $\Omega
^{(0)}(\beta\mid\mu)$ by the Legendre transformation, Eq.\ (\ref{2.9}), as
\begin{equation}
\beta F^{(0)}(\beta\mid n)=-Tr\left(  \ln\left(  1+e^{-\beta(\frac{\widehat
{p}^{2}}{2m}-\mu(\widehat{\mathbf{q}}))}\right)  -\beta\mu\left(
\widehat{\mathbf{q}}\right)  \widehat{\rho}_{e}\right)  . \label{3.6}%
\end{equation}
where use has been made of (\ref{3.2}) for the density. This result can be
expressed in terms of the Fermi operator by eliminating $\mu\left(
\widehat{\mathbf{q}}\right)  $ using (\ref{3.3}) to get
\begin{equation}
\beta F^{(0)}(\beta\mid n)=Tr\left[  \left(  1-\widehat{\rho}_{e}\right)
\ln\left(  1-\widehat{\rho}_{e}\right)  +\widehat{\rho}_{e}\ln\widehat{\rho
}_{e}+\beta\frac{\widehat{p}^{2}}{2m}\widehat{\rho}_{e}\right]  \,.
\label{3.7}%
\end{equation}
Finally, entropy and internal energy functionals can be identified from the
definition
\begin{equation}
F^{(0)}=U^{(0)}-TS^{(0)}\,, \label{3.8}%
\end{equation}
to get
\begin{equation}
U^{(0)}(\beta\mid n)=Tr\frac{\widehat{p}^{2}}{2m}\widehat{\rho}_{e}
\label{3.9}%
\end{equation}%
\begin{equation}
S^{(0)}(\beta\mid n)=-k_{B}Tr\left[  \left(  1-\widehat{\rho}_{e}\right)
\ln\left(  1-\widehat{\rho}_{e}\right)  +\widehat{\rho}_{e}\ln\widehat{\rho
}_{e}\right]  \,. \label{3.10}%
\end{equation}
Since the spectrum of $\widehat{\rho}_{e}$ is real and in the interval
$\left(  0,1\right)  $, the energy and entropy are positive. Observe that all
functional dependence upon the density occurs through the Fermi operator in
the form of Eq.\ (\ref{3.5}).

This development delineates the essential task, which is to expose these
definitions as explicit functionals of the density. Consider, for example, the
determination of $\mu^{(0)}(\mathbf{q},\beta\mid n)$ from inversion of the
density equation (\ref{3.2}). Evaluating the trace in coordinate
representation gives
\begin{equation}
n(\mathbf{r})=\left\langle \mathbf{r}\left\vert \left(  e^{\beta
(\frac{\widehat{p}^{2}}{2m}-\mu^{(0)}(\widehat{\mathbf{q}},\beta\mid
n))}+1\right)  ^{-1}\right\vert \mathbf{r}\right\rangle \, . \label{3.11}%
\end{equation}
This expression is deceptively simple, since extraction of $\mu^{(0)}$ is
complicated by the non-commutation of $\mu^{(0)}(\widehat{\mathbf{q}}%
,\beta\mid n)$ with the kinetic energy. A common \emph{approximation }is the
Thomas-Fermi form, which is obtained by replacing the operator $\mu
^{(0)}(\widehat{\mathbf{q}},\beta\mid n)$ by the c-number $\mu^{(0)}%
(\mathbf{r},\beta\mid n)$, so that the Fermi operator becomes diagonal in
momentum representation
\begin{equation}
n(\mathbf{r})\rightarrow h^{-3}\int d\mathbf{p}\left(  e^{\beta(\frac{p^{2}%
}{2m}-\mu^{(0)}(\mathbf{r},\beta\mid n))}+1\right)  ^{-1} \, . \label{3.12}%
\end{equation}
In this case, the inversion is the same as that for the uniform ideal Fermi
gas noted above $\mu^{(0)}(\mathbf{r},\beta\mid n)\rightarrow\mu_{e}%
^{(0)}(\beta,n(\mathbf{r}))$. This inversion is exact in the uniform limit,
but conditions for its validity otherwise are not known. In the following
discussion, several exact properties of the density functionals are exposed to
provide guidance and constraints on approximations for their explicit forms.

\subsection{Functionals of the statistical operator}

To put the equilibrium functionals of the last section in context, it is
useful to define related functionals of trace class operators $\widehat{\rho}$
which are Hermitian, positive, and have spectrum on $(0,1)$ but otherwise are
arbitrary. For them, define
\begin{equation}
\beta\mathcal{F}^{(0)}(\beta\mid\widehat{\rho})\equiv Tr\left[  \left(
1-\widehat{\rho}\right)  \ln\left(  1-\widehat{\rho}\right)  +\widehat{\rho
}\ln\widehat{\rho}+\beta\frac{\widehat{p}^{2}}{2m}\widehat{\rho}\right]  \,,
\label{3.13}%
\end{equation}%
\begin{equation}
\mathcal{U}^{(0)}\left[  \widehat{\rho}\right]  =Tr\frac{\widehat{p}^{2}}%
{2m}\widehat{\rho}\,, \label{3.14}%
\end{equation}%
\begin{equation}
\mathcal{S}^{(0)}\left[  \widehat{\rho}\right]  =-k_{B}Tr\left[  \left(
1-\widehat{\rho}\right)  \ln\left(  1-\widehat{\rho}\right)  +\widehat{\rho
}\ln\widehat{\rho}\right]  \,. \label{3.15}%
\end{equation}
Clearly these functionals become the density functionals of the preceding
section when evaluated at the Fermi operator, $\widehat{\rho}=\widehat{\rho
}_{e}$. More generally, $\mathcal{F}^{(0)}(\beta\mid\widehat{\rho})$ is a
convex functional of $\widehat{\rho}$ while $\mathcal{S}^{(0)}\left[
\widehat{\rho}\right]  $ is positive and concave. Furthermore, the minimum of
$\mathcal{F}^{(0)}(\beta\mid\widehat{\rho})$ occurs for $\widehat{\rho
}=\widehat{\rho}_{e}$, among all $\widehat{\rho}$ with the same density
\begin{equation}
n(\mathbf{r})=Tr\delta(\widehat{\mathbf{q}}-\mathbf{r})\widehat{\rho}\,.
\label{3.16}%
\end{equation}
To see this evaluate
\begin{equation}
\delta\left(  \mathcal{F}^{(0)}(\beta\mid\widehat{\rho})+\int d\mathbf{r}%
\lambda\left(  \mathbf{r}\right)  \left(  n(\mathbf{r})-Tr\delta
(\widehat{\mathbf{q}}-\mathbf{r})\widehat{\rho}\right)  \right)  =0\,,
\label{3.17}%
\end{equation}
where $\lambda\left(  \mathbf{r}\right)  $ is a Lagrange multiplier function.
The solution is
\begin{equation}
\widehat{\rho}_{m}=\left(  e^{\beta\left(  \frac{\widehat{p}^{2}}{2m}%
+\lambda\left(  \widehat{\mathbf{q}}\right)  \right)  }+1\right)
^{-1}=\widehat{\rho}_{e}\,. \label{3.18}%
\end{equation}
The last equality follows since $\lambda\left(  \widehat{\mathbf{q}}\right)  $
must be determined from Eq.\ (\ref{3.16}), which is the same as (\ref{3.2}).
In summary,
\begin{equation}
F^{(0)}(\beta\mid n)=\mathcal{F}^{(0)}(\beta\mid\widehat{\rho}_{e}%
)\leq\mathcal{F}^{(0)}(\beta\mid\widehat{\rho}). \label{3.19}%
\end{equation}
This result will be used below to establish upper bounds on the free energy
density functional.

\section{Scaling relations, bounds, and inequalities}

\label{sec4}

\subsection{Scaling from dimensional analysis}

In this section scaling properties of the functionals are obtained from
dimensional analysis. An alternative constructive derivation of scaling
properties is given in Appendix \ref{apB}. It is closer in formulation to the
uniform scaling widely used in ground-state DFT \cite{LevyPerdew85} but does
not expose the simplicity and profound connection with thermodynamics that
comes from the dimensional approach.

All the density dependence of the density functionals occurs through the
dependence of the Fermi operator on $\mu^{(0)}(\mathbf{r,}\beta\mid n)$ or,
equivalently, the dimensionless local ``activity''
\begin{equation}
\nu^{(0)}(\mathbf{r,}\beta\mid n)\equiv\beta\mu^{(0)}(\mathbf{r,}\beta\mid n)
\, . \label{4.1}%
\end{equation}
Define the dimensionless momentum and position operators
\begin{equation}
\widehat{y}\equiv\sqrt{\frac{\beta}{2m}}\widehat{p},\hspace{0.2in}%
\mathbf{r}^{\ast}=\frac{\mathbf{r}}{\lambda} \, , \label{4.2}%
\end{equation}
where $\lambda$ is the thermal de Broglie wavelength,
\begin{equation}
\lambda=\left(  \frac{2\pi\beta\hbar^{2}}{m}\right)  ^{1/2} \, . \label{4.2b}%
\end{equation}
Then Eq.\ (\ref{3.11}) becomes
\begin{equation}
n^{\ast}(\mathbf{r}^{\ast})\equiv\lambda^{3} n(\mathbf{r})=\left\langle
\mathbf{r}^{\ast}\left\vert \left(  e^{\widehat{y}^{2}-\nu(\widehat
{\mathbf{r}})}+1\right)  ^{-1}\right\vert \mathbf{r}^{\ast}\right\rangle
\label{4.3}%
\end{equation}
This shows that $\nu(\widehat{\mathbf{r}})$ is a functional of the
dimensionless density $n^{\ast}(\mathbf{r}^{\ast})$ alone%
\begin{equation}
\nu(\mathbf{r})=\nu^{(0)}(\mathbf{r,}\beta\mid n)=\nu^{(0)}(\mathbf{r}^{\ast
}\mid n^{\ast}). \label{4.4}%
\end{equation}
The immediate consequence of (\ref{4.4}) is that the Fermi operator and all
functionals in dimensionless form are also functionals of $n^{\ast}$ only, to
wit
\begin{equation}
\beta F^{(0)}(\beta\mid n)=F^{\ast}[n^{\ast}],\hspace{0.2in}\beta
U^{(0)}(\beta\mid n)=U^{\ast}[n^{\ast}],\hspace{0.2in}\frac{1}{k_{B}}%
S^{(0)}(\beta\mid n)=S^{\ast}[n^{\ast}]\, . \label{4.5}%
\end{equation}

Consider now the scaling transformation
\begin{equation}
\mathbf{r\rightarrow}\gamma\mathbf{r,\hspace{0.2in}}n(\mathbf{r}%
)d\mathbf{r}\rightarrow\gamma^{3}n(\gamma\mathbf{r})dr,\mathbf{\hspace{0.2in}}
V \rightarrow\gamma^{-3} V \label{4.6}%
\end{equation}
where $V$ is the volume. This transformation preserves the total number of
particles
\begin{equation}
\int_{V}n(\mathbf{r})d\mathbf{r}=N=\int_{\gamma^{-3}V}\gamma^{3}%
n(\gamma\mathbf{r})d\mathbf{r} \, . \label{4.7}%
\end{equation}
The dimensionless density then transforms according to
\begin{equation}
n^{\ast}(\mathbf{r}^{\ast})\equiv\lambda^{3} n(\mathbf{r}) \rightarrow
\gamma^{3}\lambda^{3} n(\gamma\mathbf{r})= \left(  \gamma\lambda\right)  ^{3}
n(\gamma\lambda\mathbf{r}^{\ast}) \, . \label{4.8}%
\end{equation}
This is the expected result, namely that the length scale $\lambda$ used to
define the dimensionless density itself has been scaled by $\gamma$. However,
this scaling can be canceled by a suitable scaling of the temperature in the
de Broglie wavelength, to wit
\begin{equation}
\gamma\lambda\left(  \gamma^{-2}\beta\right)  =\lambda\left(  \beta\right)  \,
. \label{4.9}%
\end{equation}
Thus, a combined scaling transformation of lengths \emph{and} of the
temperature leaves the dimensionless density invariant, and hence all the
functionals of (\ref{4.5}), unchanged. To be explicit, define the transformed
density by
\begin{equation}
n_{\gamma}(\mathbf{r})\equiv\gamma^{3}n(\gamma\mathbf{r}) \, . \label{4.10}%
\end{equation}
Then from (\ref{4.5}) and the invariance of $n^{\ast}$, the desired scaling
properties follow:
\begin{equation}
F^{(0)}(\beta\mid n)=\gamma^{-2}F^{(0)}(\gamma^{-2}\beta\mid n_{\gamma
}),\hspace{0.2in}U^{(0)}(\beta\mid n)=\gamma^{-2}U^{(0)}(\gamma^{-2}\beta\mid
n_{\gamma}) \label{4.11}%
\end{equation}%
\begin{equation}
S^{(0)}(\beta\mid n)=S^{(0)}(\gamma^{-2}\beta\mid n_{\gamma}) \, .
\label{4.12}%
\end{equation}

\subsection{Bounds and inequalities}

In this section specific upper and lower bounds are described. In addition,
the scaling laws are used to obtain inequalities for the various functionals
for different values of their arguments (temperature and density).

\subsubsection{Lower bound}

It is well-known \cite{SearsParrDinur80,Harriman85,Herring86} that the
non-interacting kinetic energy density functional at zero temperature is
bounded from below by the von Weizs\"acker functional
\begin{equation}
U_{vW}[n]\equiv\frac{\hbar^{2}}{8m}\int d\mathbf{r}\frac{\left\vert \nabla
n(\mathbf{r})\right\vert ^{2}}{n(\mathbf{r})}=\frac{\hbar^{2}}{2m}\int
d\mathbf{r}\left\vert \nabla\varphi(\mathbf{r})\right\vert ^{2} \, .
\label{4.13}%
\end{equation}
The last equality is obtained by introducing the square root of the density (a
positive function)
\begin{equation}
n(\mathbf{r})=\varphi(\mathbf{r})^{2} \, . \label{4.14}%
\end{equation}
We now show that the von Weizs\"acker functional is a lower bound for 
the kinetic energy density functional at finite temperature as well.

Since $\widehat{\rho}_{e}$ is a Hermitian, positive operator it has a square
root:
\begin{equation}
\widehat\rho_{e}=(\widehat\psi)^{2} \, . \label{4.15}%
\end{equation}
Then $U^{(0)}(\beta\mid n)$ from Eq.\ (\ref{3.9}) can be written
\begin{equation}
U^{(0)}(\beta\mid n)=\frac{1}{2m}Tr\left\vert \widehat{p}\widehat
\psi\right\vert ^{2}=\frac{\hbar^{2}}{2m}\int d\mathbf{r}d\mathbf{r}^{\prime
}\left\vert \nabla\left\langle \mathbf{r}\right\vert \widehat\psi\left\vert
\mathbf{r}^{\prime}\right\rangle \right\vert ^{2} \, . \label{4.16}%
\end{equation}
Note the similarity of this exact result to the approximate von Weizs\"acker
result (\ref{4.13}) (the relationship of $\widehat{\psi}$ to the square root
of the density is $\varphi(\mathbf{r})=\sqrt{\left\langle \mathbf{r}%
\right\vert \widehat\psi^{2}\left\vert \mathbf{r}\right\rangle }$). To extract
the von Weizs\"acker contribution, define
\begin{equation}
X\left(  \mathbf{r},\mathbf{r}^{\prime}\right)  \equiv\varphi^{-1}%
(\mathbf{r})\left\langle \mathbf{r}\right\vert \widehat\psi\left\vert
\mathbf{r}^{\prime}\right\rangle \, . \label{4.17}%
\end{equation}
Then
\begin{align}
U^{(0)}(\beta &  \mid n)=\frac{\hbar^{2}}{2m}\int d\mathbf{r}\left\vert
\nabla\varphi(\mathbf{r})\right\vert ^{2}\int d\mathbf{r}^{\prime\prime
}\left\vert X\left(  \mathbf{r},\mathbf{r}^{\prime\prime}\right)  \right\vert
^{2}+\frac{\hbar^{2}}{2m}\int d\mathbf{r}\left\vert \varphi(\mathbf{r}%
)\right\vert ^{2}\int d\mathbf{r}^{\prime\prime}\left\vert \nabla X\left(
\mathbf{r},\mathbf{r}^{\prime\prime}\right)  \right\vert ^{2}\nonumber\\
&  +\frac{\hbar^{2}}{2m}\int d\mathbf{r}\varphi(\mathbf{r})\nabla
\varphi(\mathbf{r})\nabla\int d\mathbf{r}^{\prime\prime}\left\vert X\left(
\mathbf{r},\mathbf{r}^{\prime\prime}\right)  \right\vert ^{2} \, .
\label{4.18}%
\end{align}
But
\begin{equation}
\int d\mathbf{r}^{\prime\prime}\left\vert X\left(  \mathbf{r},\mathbf{r}%
^{\prime\prime}\right)  \right\vert ^{2}=\frac{1}{n(\mathbf{r})}\int
d\mathbf{r}^{\prime\prime}\left\langle \mathbf{r}\right\vert \widehat{\psi
}\left\vert \mathbf{r}^{\prime\prime}\right\rangle \left\langle \mathbf{r}%
^{\prime\prime}\right\vert \widehat{\psi}\left\vert \mathbf{r}\right\rangle =1
\, , \label{4.19}%
\end{equation}
so
\begin{align}
U^{(0)}(\beta &  \mid n)=\frac{\hbar^{2}}{2m}\int d\mathbf{r}\left\vert
\nabla\varphi(\mathbf{r})\right\vert ^{2}+\frac{\hbar^{2}}{2m}\int
d\mathbf{r}\left\vert \varphi(\mathbf{r})\right\vert ^{2}\int d\mathbf{r}%
^{\prime\prime}\left\vert \nabla X\left(  \mathbf{r},\mathbf{r}^{\prime\prime
}\right)  \right\vert ^{2}\nonumber\\
&  \geq U_{vW}\left[  n\right]  \label{4.20}%
\end{align}
This is the desired result, showing that $U_{vW}$ is a lower bound for 
$U^{(0)}$.
The analogue in ground-state DFT is the so-called Pauli-term decomposition
\cite{TalBader78,BartolottiAcharya82,Harriman87,LevyOu-Yang88} of the KS
kinetic energy
\begin{equation}
{\mathcal{T}}_{s} = {\mathcal{T}}_{vW} + {\mathcal{T}}_{\theta}\,\, ,
\hspace*{0.5in} {\mathcal{T}}_{\theta}\ge0 \, . \label{4.20A}%
\end{equation}
Eq.\ (\ref{4.20}) also provides a bound for the free
energy functional, namely
\begin{equation}
F(\beta\mid n)\geq U_{vW}\left[  n\right]  -TS(\beta\mid n) \, . \label{4.21}%
\end{equation}

In obtaining (\ref{4.20}), no use has been made of the integral over
$\mathbf{r}$, so provided that no traceless forms are added to the energy
densities, the bound holds for them too:
\begin{equation}
F(\beta\mid n)\equiv\int d\mathbf{r}f(\mathbf{r},\beta\mid n),\hspace
{0.2in}U(\beta\mid n)\equiv\int d\mathbf{r}u(\mathbf{r},\beta\mid
n),\hspace{0.2in}S(\beta\mid n)\equiv\int d\mathbf{r}s(\mathbf{r},\beta\mid n)
\label{4.22}%
\end{equation}
\begin{equation}
u(\mathbf{r},\beta\mid n)=u_{vW}(\mathbf{r}\mid n)+\frac{\hbar^{2}}%
{2m}n(\mathbf{r})\int d\mathbf{r}^{\prime\prime}\left\vert \nabla X\left(
\mathbf{r},\mathbf{r}^{\prime\prime}\right)  \right\vert ^{2}\geq
u_{vW}(\mathbf{r}\mid n) \label{4.23}%
\end{equation}
\begin{equation}
f(\mathbf{r},\beta\mid n)\geq u_{vW}(\mathbf{r}\mid n)-Ts(\mathbf{r},
\beta\mid n) \, . \label{4.24}%
\end{equation}

\subsubsection{Upper bound}

Upper bounds are provided by (\ref{3.19}) if a suitable choice for
$\widehat{\rho}$ can be found. In practice, it is common to use the local
density approximation, which in this context leads to the Thomas-Fermi form
for the density.
The question arises as to what statistical operator corresponds to the
Thomas-Fermi approximation. More specifically, what $\widehat{\rho}$ has the
Thomas-Fermi form for the density? The necessary condition is
\begin{equation}
\left\langle \mathbf{r}\left\vert \widehat{\rho}_{TF}\right\vert
\mathbf{r}\right\rangle =h^{-3}\int d\mathbf{p}\left(  e^{\beta(\frac{p^{2}%
}{2m}-\mu_{e}(\beta,n\left(  \mathbf{r}\right)  ))}+1\right)  ^{-1}\,.
\label{4.25}%
\end{equation}
However, this condition determines only the diagonal matrix elements of
$\widehat{\rho}_{TF}$. To fix the statistical operator definitively, consider
its associated Wigner distribution $n_{TF}\left(  p,\mathbf{R}\right)  $,
defined by
\begin{equation}
\left\langle \mathbf{r}\right\vert \widehat{\rho}_{TF}\left\vert
\mathbf{r}^{\prime}\right\rangle =\frac{1}{h^{3}}\int d\mathbf{p}e^{\frac
{i}{\hbar}\mathbf{p\cdot q}}n_{TF}\left(  p,\mathbf{R}\right)  ,\hspace
{0.25in}\mathbf{R}=\frac{\mathbf{r+r}^{\prime}}{2},\hspace{0.25in}%
\mathbf{q}=\mathbf{r-r}^{\prime}\,. \label{4.26}%
\end{equation}
Specialization to $\mathbf{q=0}$ and comparison with (\ref{4.25}) determines
$n_{TF}\left(  p,\mathbf{R}\right)  $ as
\begin{equation}
n_{TF}\left(  p,\mathbf{R}\right)  =\left(  e^{\beta(\frac{p^{2}}{2m}-\mu
_{e}(\beta,n\left(  \mathbf{r}\right)  ))}+1\right)  ^{-1}\,. \label{4.28}%
\end{equation}
With the statistical operator determined, an upper bound is given by
(\ref{3.19}) with $\rho=\rho_{TF}$%
\begin{equation}
F^{(0)}(\beta\mid n)\leq\mathcal{F}^{(0)}(\beta\mid\widehat{\rho}%
_{TF})=\mathcal{U}^{(0)}[\widehat{\rho}_{TF}]-T\mathcal{S}^{(0)}[\widehat
{\rho}_{TF}]. \label{4.29}%
\end{equation}

The calculation of $\mathcal{U}^{(0)}[\rho_{TF}]$ is straightforward
\begin{align}
\mathcal{U}^{(0)}[\widehat{\rho}_{TF}]  &  =-\frac{\hbar^{2}}{2m}\int
d\mathbf{r}\left[  \nabla^{\prime2}\left\langle \mathbf{r}^{\prime}\right\vert
\widehat{\rho}_{TF}\left\vert \mathbf{r}\right\rangle \right]  _{\mathbf{r}%
^{\prime}=\mathbf{r}}=\int d\mathbf{r}\frac{1}{h^{3}}\int d\mathbf{p}%
\frac{\mathbf{p}^{2}}{2m}n_{TF}\left(  p,\mathbf{r}\right) \nonumber\\
&  =\int d\mathbf{r}u_{TF}(\beta,n\left(  \mathbf{r}\right)  )\,, \label{4.30}%
\end{align}
where $u_{TF}(\beta,n\left(  \mathbf{r}\right)  )$ is the finite temperature
Thomas-Fermi energy density
\begin{equation}
u_{TF}(\beta,n\left(  \mathbf{r}\right)  )=\frac{1}{h^{3}}\int d\mathbf{p}%
\frac{\mathbf{p}^{2}}{2m}\left(  e^{\beta(\frac{p^{2}}{2m}-\mu_{e}%
(\beta,n\left(  \mathbf{r}\right)  ))}+1\right)  ^{-1}\,. \label{4.31}%
\end{equation}
To complete the calculation of the upper bound it is necessary to evaluate
\begin{equation}
\mathcal{S}^{(0)}\left[  \widehat{\rho}_{TF}\right]  =-k_{B}\int
d\mathbf{r}\left\langle \mathbf{r}\left\vert \left[  \left(  1-\widehat{\rho
}_{TF}\right)  \ln\left(  1-\widehat{\rho}_{TF}\right)  +\widehat{\rho}%
_{TF}\ln\widehat{\rho}_{TF}\right]  \right\vert \mathbf{r}\right\rangle \,.
\label{4.32}%
\end{equation}
Although the density dependence of all matrix elements of $\widehat{\rho}%
_{TF}$ is explicit, the evaluation of non-linear functions of this operator is
still a difficult problem and the explicit form for (\ref{4.32}) has not yet
been determined.

In any event, the upper bound for the free energy density functional is
\begin{align}
F^{(0)}(\beta &  \mid n)\leq\int d\mathbf{r}u_{TF}(\beta,n\left(
\mathbf{r}\right)  )\nonumber\\
&  +\beta^{-1}\int d\mathbf{r}\left\langle \mathbf{r}\left\vert \left[
\left(  1-\widehat{\rho}_{TF}\right)  \ln\left(  1-\widehat{\rho}_{TF}\right)
+\widehat{\rho}_{TF}\ln\widehat{\rho}_{TF}\right]  \right\vert \mathbf{r}%
\right\rangle \,. \label{4.33}%
\end{align}

\subsubsection{Inequalities}

A class of inequalities can be obtained from Eq.\ (\ref{3.19}) and the scaling
laws. Recall the definition of $n_{\gamma}$ in (\ref{4.10}), and choose
$\widehat{\rho}=\widehat{\rho}_{e}(\gamma^{-2}\beta\mid n_{\gamma})$ in
(\ref{3.19}),
\begin{equation}
F^{(0)}\left(  \beta\mid n_{\gamma}\right)  =\mathcal{F}^{(0)}\left(  \beta %
\mid\widehat{\rho}_{e}(\beta\mid n_{\gamma})\right)  \leq\mathcal{F}^{(0)} %
\left(\beta\mid\widehat{\rho}_{e}(\gamma^{-2}\beta\mid n_{\gamma})\right)  \,,
\label{4.35}%
\end{equation}
and then use the definition of $\mathcal{F}^{(0)}\left( \beta\mid\rho\right) $ 
to get
\begin{equation}
F^{(0)}\left(  \beta\mid n_{\gamma}\right)  -F^{(0)}\left(  \gamma^{-2}%
\beta\mid n_{\gamma}\right)  \leq\left(  \gamma^{2}-1\right)  TS^{(0)}%
(\gamma^{-2}\beta\mid n_{\gamma})\,. \label{4.36}%
\end{equation}
Use of the scaling laws (\ref{4.11},\ref{4.12}) to convert functions 
of $\gamma^{-2}\beta$ to
corresponding functions of $\beta$ leads to
\begin{equation}
F^{(0)}\left(  \beta\mid n_{\gamma}\right)  -\gamma^{2}F\left(  \beta\mid
n\right)  \leq\left(  \gamma^{2}-1\right)  TS^{(0)}\left(  \beta\mid n\right)
\,. \label{4.36a}%
\end{equation}
Since the entropy is non-negative, an inequality for constant temperature and
different densities is
\begin{equation}
F^{(0)}\left(  \beta\mid n_{\gamma}\right)  \leq\gamma^{2}F^{(0)}\left(
\beta\mid n\right)  ,\hspace{0.2in}\gamma\leq1\,. \label{4.38}%
\end{equation}
Use of the scaling laws once more to convert $F^{(0)}\left(  \beta\mid
n_{\gamma}\right)  $ to $\gamma^{2}F^{(0)}\left(  \gamma^{2}\beta\mid
n\right)  $ gives the inequality for constant density
\begin{equation}
F^{(0)}\left(  \gamma^{2}\beta\mid n\right)  \leq F^{(0)}\left(  \beta\mid
n\right)  ,\hspace{0.2in}\gamma\leq1\,. \label{4.39}%
\end{equation}

Repetition of this analysis for the choice
\begin{equation}
F^{(0)}\left(  \gamma^{-2}\beta\mid n_{\gamma}\right)  =\mathcal{F}^{(0)}\left(
\gamma^{-2}\beta\mid\widehat{\rho}_{e}(\gamma^{-2}\beta\mid n_{\gamma
})\right)  \leq\mathcal{F}^{(0)}\left(  \gamma^{-2}\beta\mid\widehat{\rho}_{e}%
(\beta\mid n_{\gamma})\right)  \,, \label{4.40}%
\end{equation}
leads to
\begin{equation}
F^{(0)}\left(  \gamma^{-2}\beta\mid n_{\gamma}\right)  -F^{(0)}\left(
\beta\mid n_{\gamma}\right)  \leq\left(  1-\gamma^{2}\right)  TS^{(0)}%
(\beta\mid n_{\gamma})\,, \label{4.41}%
\end{equation}
and scaling gives the reverse inequalities for $\gamma\geq1$:
\begin{equation}
\gamma^{2}F^{(0)}\left(  \beta\mid n\right)  \leq F^{(0)}\left(  \beta\mid
n_{\gamma}\right)  ,\hspace{0.2in}\gamma\geq1\,. \label{4.42}%
\end{equation}
In consequence, it also is true that
\begin{equation}
F^{(0)}\left(  \beta\mid n\right)  \leq F^{(0)}\left(  \gamma^{2}\beta\mid
n\right)  ,\hspace{0.2in}\gamma\geq1\,. \label{4.43}%
\end{equation}

Appendix \ref{apC} has further consideration of Eqs.\ (\ref{4.36}) and
(\ref{4.41}) to reach corresponding inequalities for the entropy and energy
functionals. The results are summarized in the Table. These inequalities imply
that the entropy and energy are monotonically increasing with $T$, while the
free energy is monotonically decreasing with $T$, at constant $n $ .
\vspace*{0.25in}%

\begin{tabular}
[c]{|c|c|}\hline
Constant $n$ & Constant $\beta$\\\hline
$F^{(0)}\left(  \beta\mid n\right)  \lesseqgtr F^{(0)}\left(  \gamma^{2}%
\beta\mid n\right)  ,\hspace{0.2in}\gamma\gtreqless1$ & $F^{(0)}\left(
\beta\mid n_{\gamma}\right)  \gtreqless\gamma^{2}F^{(0)}\left(  \beta\mid
n\right)  ,\hspace{0.2in}\gamma\gtreqless1$\\\hline
$U^{(0)}\left(  \beta\mid n\right)  \gtreqless U^{(0)}\left(  \gamma^{2}%
\beta\mid n\right)  ,\hspace{0.2in}\gamma\gtreqless1$ & $U^{(0)}\left(
\beta\mid n_{\gamma}\right)  \lesseqgtr\gamma^{2}U^{(0)}\left(  \beta\mid
n\right)  ,\hspace{0.2in}\gamma\gtreqless1$\\\hline
$S^{(0)}(\beta\mid n)\gtreqless S^{(0)}(\gamma^{2}\beta\mid n),\hspace
{0.2in}\gamma\gtreqless1$ & $S^{(0)}(\beta\mid n_{\lambda})\lesseqgtr
S^{(0)}(\beta\mid n),\hspace{0.2in}\gamma\gtreqless1$\\\hline
\end{tabular}
\vspace{0.25in}

Return to (\ref{4.36a}) and use scaling to write it as
\begin{equation}
F^{(0)}\left(  \gamma^{2}\beta\mid n\right)  -F\left(  \beta\mid n\right)
\leq\frac{\left(  \gamma^{2}-1\right)  }{\gamma^{2}}TS^{(0)}\left(  \beta\mid
n\right)  \, . \label{4.44}%
\end{equation}
Set $\gamma=1+\delta$ with $\delta\rightarrow0$ to get
\begin{equation}
\frac{dF^{(0)}\left(  \beta\mid n\right)  }{d\beta}\mid_{n}=TS^{(0)}(\beta\mid
n)\, . \label{4.55}%
\end{equation}
This is a thermodynamic identity, which confirms the fact that the functionals
considered here are all derived from the grand potential (\ref{3.1}) for a
non-uniform system at equilibrium.

\section{Discussion}

\label{sec5}The analysis of density functionals relevant to finite temperature
DFT has been placed in the context of equilibrium statistical mechanics for a
non-uniform system. The free energy functional of (\ref{2.9}) is precisely the
same as that introduced by Mermin, but obtained here as a thermodynamic change
of variables via Legendre transformation. This recognition of the
thermodynamic context for DFT is expected to play a more important role when
the functionals are parameterized by temperature, than for familiar 
ground-state DFT.   The origin of the density dependence through inversion 
of the
relationship of $n(\mathbf{r})$ to $\mu(\mathbf{r})$ allows a formal
representation of all functionals for constructive evaluation: the more
general constrained search methods are replaced by an explicit, albeit very
difficult, many-body evaluation. The analysis here is restricted to the
non-interacting functionals for which that many-body problem can be solved.
Nevertheless, the necessary inversion to find $\mu(\mathbf{r})=$ $\mu
^{(0)}(\mathbf{r},\beta\mid n)$ remains a difficult mathematical task. Hence,
construction of approximate functionals is required in this case as well. 
The objective here has been to provide both context and constraints for 
that process.

The non-interacting functionals of section \ref{sec3} are deceptively simple
and still defy exact evaluation. However, they are in a form convenient for
exposing some of their general properties. Linear coordinate scaling is an
example.  For it, simple dimensional analysis provides the exact results. The
results are analogous to those obtained for ground-state DFT, except the
temperature is an additional variable that must scale along with the density.
This is an example of a general outcome.  Many of the finite-temperature 
results conform to obvious expectations from the ground-state theory, but 
some are not predictable by such extrapolation.   

Application of these results includes determination of the volume dependence
of the functionals (see, for example, Chihara and Yamagiwa \cite{Thermo}).
Identification of the von Weizs\"{a}cker lower bound is perhaps expected, but
the proof also provides a representation for the exact positive correction
that may be suitable for further analysis. Upper bounds are provided by noting
that all functionals depend on the density through the Fermi operator, and
that the free energy is the extremum of a convex functional of constrained
single-particle reduced statistical operators $\mathcal{F}^{(0)}(\beta
\mid\widehat{\rho})$. As an application, the single-particle reduced
statistical operator corresponding to the TF approximation was constructed and
shown to yield the TF approximation for a class of functionals, but not for
the entropy in particular. Thus TF does not result from some approximate
single-particle statistical operator. This will be discussed further elsewhere.

The inequalities of the last section are a consequence of the scaling laws and
convexity of the $\mathcal{F}^{(0)}(\beta\mid\widehat{\rho})$. Since the
scaling transformations involve both temperature and density, two classes of
inequalities are described: those for functionals at different densities but
the same constant temperature, and for the temperature dependence of
functionals for constant density. Ultimately, it is expected that such
inequalities have an instructive thermodynamic interpretation (see for example
(\ref{4.55})).

\section{Acknowledgement}

\label{sec6} This work was supported under US DOE Grant DE-SC0002139. We
acknowledge informative conversations with Stefano Pittalis.

\appendix

\section{Determination of $\Omega^{(0)}(\beta\mid\mu)$}

\label{apA}

\subsection{Derivation of Eq.\ (\ref{3.1})}

Development of the thermodynamics for an ideal gas occurs in every textbook on
equilibrium quantum statistical mechanics. For completeness, the extension to
non-uniform systems to derive (\ref{3.1}) is described briefly here. First
write Eq.\ (\ref{2.7}) in the form
\begin{equation}
\Omega^{(0)}(\beta\mid\mu)=-\beta^{-1}\ln\sum_{N=0}^{\infty}Tr^{(N)}%
e^{-\beta\left(  \sum_{i=1}^{N}a(\widehat{\mathbf{q}}_{i},\widehat{\mathbf{p}%
}_{i})\right)  } \label{aq.1}%
\end{equation}
with the single-particle operator $a(\widehat{\mathbf{q}}_{i},\widehat
{\mathbf{p}}_{i})$ given by
\begin{equation}
a(\widehat{\mathbf{q}}_{i},\widehat{\mathbf{p}}_{i})=\frac{\widehat{p}_{i}%
^{2}}{2m}-\mu(\widehat{\mathbf{q}}_{i}) \label{a.2}%
\end{equation}
Since $a(\widehat{\mathbf{q}}_{i},\widehat{\mathbf{p}}_{i})$ is a Hermitian
operator, it has an associated complete basis set for the single-particle
Hilbert space
\begin{equation}
a(\widehat{\mathbf{q}}_{i},\widehat{\mathbf{p}}_{i})\mid\alpha_{i}%
>=\varepsilon_{\alpha_{i}}\mid\alpha_{i}>\,. \label{a.3}%
\end{equation}
Observe that the \textquotedblleft$\alpha$\textquotedblright\ are state labels
while the \textquotedblleft$i$\textquotedblright\ are particle indices. The
$N$ particle Hilbert space then is spanned by the anti-symmetrized product
basis $\mathcal{S}\mid\alpha_{1},..,\alpha_{N}>$, and
\begin{equation}
\sum_{i=1}^{N}a(\widehat{\mathbf{q}}_{i},\widehat{\mathbf{p}}_{i}%
)\mathcal{S}\mid\alpha_{1},..,\alpha_{N}>=\left(  \sum_{i=1}^{N}%
\varepsilon_{\alpha_{i}}\right)  \mathcal{S}\mid\alpha_{1},..,\alpha_{N}>\,.
\label{a.4}%
\end{equation}
These total energy eigenvalues can be specified in an equivalent way in terms
of the occupation numbers for a given quantum state
\begin{equation}
\sum_{i=1}^{N}\varepsilon_{\alpha_{i}}=\sum_{\alpha}\varepsilon_{\alpha
}n_{\alpha}\,. \label{a,5}%
\end{equation}
The sum on the right side is over all single-particle states, and $n_{\alpha}$
is the number of times the quantum number $\alpha$ occurs in the state
$\mid\alpha_{1},..,\alpha_{N}>$, restricted by $\sum_{\alpha}n_{\alpha}=N$.

Since the states are anti-symmetrized, there is no meaning to assignment of
which particles are in the state with quantum number $\alpha$, only the total
number of particles in that state $n_{\alpha}$. Consequently, the states can
be labeled by $\left\{  n_{\alpha}\right\}  $ instead of $\left\{  \alpha
_{i}\right\}  $
\begin{equation}
\mathcal{S}\mid\alpha_{1},..,\alpha_{N}>\longleftrightarrow\mid\left\{
n_{\alpha}\right\}  > \label{a.6}%
\end{equation}
In this occupation number representation, the effects of symmetrization are
accounted for since one set of $\left\{  n_{\alpha}\right\}  $ represents all
the states generated by permutations. The trace for the grand potential can be
carried out in this occupation number representation as follows:
\begin{align}
\Omega^{(0)}(\beta &  \mid\mu)=-\beta^{-1}\ln\sum_{N=0}^{\infty}\sum_{\left\{
n_{\alpha}\right\}  }<\left\{  n_{\alpha}\right\}  \mid e^{-\beta\left(
\sum_{i=1}^{N}a(\widehat{\mathbf{q}}_{i},\widehat{\mathbf{p}}_{i})\right)
}\mid\left\{  n_{\alpha}\right\}  >\nonumber\\
&  =-\beta^{-1}\ln\sum_{N=0}^{\infty}\prod_{\alpha}\sum_{n_{\alpha}}^{\prime
}e^{-\beta\varepsilon_{\alpha}n_{\alpha}}=-\beta^{-1}\ln\sum_{N =0}%
^{\infty}\prod_{\alpha}\sum_{n_{\alpha}}e^{-\beta\varepsilon_{\alpha}%
n_{\alpha}}\delta\left(  N,\sum_{\alpha}n_{\alpha}\right)  \label{a.7}%
\end{align}
Primes over summations over $n_{\alpha}$ indicate the restriction to
$\sum_{\alpha}n_{\alpha}=N$, a restriction that is removed in the last
equality by the explicit Kronecker delta. The summation over $N$ is then
performed first, giving
\begin{equation}
\Omega^{(0)}(\beta\mid\mu)=-\beta^{-1}\sum_{\alpha}\ln\sum_{n_{\alpha}%
}e^{-\beta\varepsilon_{\alpha}n_{\alpha}}\,. \label{a.8}%
\end{equation}

Fermion occupation numbers are all $n_{\alpha}=0,1$ so
\begin{align}
\Omega^{(0)}(\beta &  \mid\mu)=-\beta^{-1}\sum_{\alpha}\ln\left(
1+e^{-\beta\varepsilon_{\alpha}}\right)  =-\beta^{-1}\sum_{\alpha}\left\langle
\alpha_{1}\right\vert \ln\left(  1+e^{-\beta a(\widehat{\mathbf{q}}%
_{1},\widehat{\mathbf{p}}_{1})}\right)  \mid\alpha_{1}>\nonumber\\
&  =-\beta^{-1}Tr\ln\left(  1+e^{-\beta a(\widehat{\mathbf{q}}_{1}%
,\widehat{\mathbf{p}}_{1})}\right)  \label{a.9}%
\end{align}
This is the result quoted in the last line of (\ref{3.1}).

\subsection{Proof of $n(\mathbf{r})=-\delta\Omega^{(0)}(\beta\mid\mu
)/\delta\mu\left(  \mathbf{r}\right)  \mid_{\beta}$}

The functional derivatives entail use of some operator algebra. To illustrate
this, the derivation of Eq.\ (\ref{3.2}) from (\ref{3.1}) is described. Begin
with the functional derivative of (\ref{3.1}),
\begin{equation}
-\frac{\delta\Omega^{(0)}(\beta\mid\mu)}{\delta\mu\left(  \mathbf{r}\right)
}\Big\vert_{\beta}=\beta^{-1}Tr^{(1)}\frac{\delta}{\delta\mu\left(
\mathbf{r}\right)  }\ln\left(  1+e^{-\beta\left(  \frac{\widehat{p}^{2}}%
{2m}-\mu(\widehat{\mathbf{q}})\right)  }\right)  \, . \label{a.10}%
\end{equation}
To calculate the derivative of the logarithm a useful representation is
\begin{equation}
\ln(1+\widehat{X})=\int_{0}^{1}d\lambda\partial_{\lambda}\ln(1+\lambda
\widehat{X})=\int_{0}^{1}d\lambda(1+\lambda\widehat{X})^{-1}\widehat{X} \, ,
\label{a.11}%
\end{equation}
where
\begin{equation}
\widehat{X}=e^{-\beta\left(  \frac{\widehat{p}^{2}}{2m}-\mu(\widehat
{\mathbf{r}}\mathbf{,}\beta\mid n)\right)  } \, . \label{a.12}%
\end{equation}
The functional derivative is then
\begin{equation}
\frac{\delta}{\delta n(\mathbf{r})}\ln(1+\widehat{X})=\int_{0}^{1}%
d\lambda(1+\lambda\widehat{X})^{-1}\frac{\delta\widehat{X}}{\delta
n(\mathbf{r})}(1+\lambda\widehat{X})^{-1} \label{a.13}%
\end{equation}
In the present context, this expression occurs under a trace, so use of the
cyclic invariance of the trace gives
\begin{align}
-\frac{\delta\Omega^{(0)}(\beta\mid\mu)}{\delta\mu\left(  \mathbf{r}\right)
}  &  \Big\vert_{\beta}=\beta^{-1}Tr\int_{0}^{1}d\lambda(1+\lambda\widehat
{X})^{-2}\frac{\delta\widehat{X}}{\delta n(\mathbf{r})}\nonumber\\
&  =\beta^{-1}Tr(1+\widehat{X})^{-1}\frac{\delta\widehat{X}}{\delta
n(\mathbf{r})} \label{a.14}%
\end{align}

Next, the derivative of the exponential (\ref{a.12}) can be defined as
\begin{equation}
\frac{\delta e^{-\beta\left(  \frac{\widehat{p}^{2}}{2m}-\mu(\widehat
{\mathbf{r}})\right)  }}{\delta\mu(\mathbf{r})}\equiv\lim\frac{e^{-\beta
\left(  \frac{\widehat{p}^{2}}{2m}-\mu(\widehat{\mathbf{r}})-\delta\mu\right)
}-e^{-\beta\left(  \frac{\widehat{p}^{2}}{2m}-\mu(\widehat{\mathbf{r}%
})\right)  }}{\delta\mu(\mathbf{r})} \, . \label{a.15}%
\end{equation}
Then writing
\begin{equation}
e^{-\beta\left(  \frac{\widehat{p}^{2}}{2m}-\mu(\widehat{\mathbf{r}}%
)-\delta\mu(\widehat{\mathbf{r}})\right)  }\equiv e^{-x\left(  \frac
{\widehat{p}^{2}}{2m}-\mu(\widehat{\mathbf{r}})-\delta\mu(\widehat{\mathbf{r}%
})\right)  }\mid_{x=\beta} \label{a.16}%
\end{equation}
\[
e^{-x\left(  \frac{\widehat{p}^{2}}{2m}-\mu(\widehat{\mathbf{r}})-\delta
\mu\right)  }=e^{-x\left(  \frac{\widehat{p}^{2}}{2m}-\mu(\widehat{\mathbf{r}%
})\right)  }U(x,\delta\mu),
\]
an equation for $U(x,\delta\mu)$ \ is obtained:
\begin{equation}
\partial_{x}U(x,\delta\mu)=e^{x\left(  \frac{\widehat{p}^{2}}{2m}-\mu
(\widehat{\mathbf{r}})\right)  }\delta\mu(\widehat{\mathbf{r}})e^{-x\left(
\frac{\widehat{p}^{2}}{2m}-\mu(\widehat{\mathbf{r}})\right)  }U(x,\delta\mu)
\, . \label{a.17}%
\end{equation}
Integrating this equation gives
\begin{align}
U(x,\delta\mu)  &  =1+\int_{0}^{x}dx^{\prime}e^{x^{\prime}\left(
\frac{\widehat{p}^{2}}{2m}-\mu(\widehat{\mathbf{r}})\right)  }\delta\mu
e^{-x^{\prime}\left(  \frac{\widehat{p}^{2}}{2m}-\mu(\widehat{\mathbf{r}%
})\right)  }U(x^{\prime},\delta\mu)\nonumber\\
&  =1+\int_{0}^{x}dx^{\prime}e^{x^{\prime}\left(  \frac{\widehat{p}^{2}}%
{2m}-\mu(\widehat{\mathbf{r}})\right)  }\delta\mu e^{-x^{\prime}\left(
\frac{\widehat{p}^{2}}{2m}-\mu(\widehat{\mathbf{r}})\right)  }+\text{ order
}\left(  \delta\mu\right)  ^{2} ], , \label{a.18}%
\end{align}
and so
\begin{equation}
e^{-\beta\left(  \frac{\widehat{p}^{2}}{2m}-\mu-\delta\mu\right)  }%
=e^{-\beta\left(  \frac{\widehat{p}^{2}}{2m}-\mu\right)  }+\int_{0}^{\beta
}dx^{\prime}e^{\left(  x^{\prime}-\beta\right)  \left(  \frac{\widehat{p}^{2}%
}{2m}-\mu\right)  }\delta\mu e^{-x^{\prime}\left(  \frac{\widehat{p}^{2}}%
{2m}-\mu\right)  }+\text{ order }\left(  \delta\mu\right)  ^{2} \, .
\label{a.19}%
\end{equation}
The derivative is then
\begin{equation}
\frac{\delta e^{-\beta\left(  \frac{\widehat{p}^{2}}{2m}-\mu(\widehat
{\mathbf{r}})\right)  }}{\delta\mu(\mathbf{r})}\equiv\int_{0}^{\beta
}dx^{\prime}e^{\left(  x^{\prime}-\beta\right)  \left(  \frac{\widehat{p}^{2}%
}{2m}-\mu(\widehat{\mathbf{r}})\right)  }\delta\left(  \widehat{\mathbf{r}%
}-\mathbf{r}\right)  e^{-x^{\prime}\left(  \frac{\widehat{p}^{2}}{2m}%
-\mu(\widehat{\mathbf{r}})\right)  } \, . \label{a.20}%
\end{equation}

With these results, and again using the cyclic invariance of the trace,
(\ref{a.10}) becomes
\begin{equation}
-\frac{\delta\Omega^{(0)}(\beta\mid\mu)}{\delta\mu\left(  \mathbf{r}\right)
}\Big\vert_{\beta}=\beta^{-1}Tr(e^{\beta\left(  \frac{\widehat{p}^{2}}{2m}%
-\mu(\widehat{\mathbf{r}})\right)  }+1)^{-1}\delta\left(  \widehat{\mathbf{r}%
}-\mathbf{r}\right)  =n(\mathbf{r}) \label{a.21}%
\end{equation}
which is Eq.\ (\ref{3.2}) of the text.

\section{Scaling from unitary transformation}

\label{apB}

Define the one-parameter family of operators
\begin{equation}
\widehat{U}(z)=e^{\frac{i}{2\hbar}z\left(  \widehat{\mathbf{q}}\mathbf{\cdot
}\widehat{\mathbf{p}}+\widehat{\mathbf{p}}\mathbf{\cdot}\widehat{\mathbf{q}%
}\right)  } \, , \label{b.1}%
\end{equation}
and the related operators
\begin{equation}
\widehat{X}(z)=\widehat{U}(z)\widehat{\mathbf{q}}\widehat{U}^{-1}%
(z),\hspace{0.25in}\widehat{Y}(z)=\widehat{U}(z)\widehat{\mathbf{p}}%
\widehat{U}^{-1}(z) \, . \label{b.2}%
\end{equation}
Then differentiation with respect to $z$
\begin{equation}
\frac{d\widehat{X(}z)}{dz}=\widehat{U}(z)\frac{i}{2\hbar}\left[  \left(
\widehat{\mathbf{q}}\mathbf{\cdot}\widehat{\mathbf{p}}+\widehat{\mathbf{p}%
}\mathbf{\cdot}\widehat{\mathbf{q}}\right)  ,\widehat{\mathbf{q}}\right]
\widehat{U}(z)^{-1}=X(z) \, , \label{b.3}%
\end{equation}
\begin{equation}
\frac{d\widehat{Y(}z)}{dz}=\widehat{U}(z)\frac{i}{2\hbar}\left[  \left(
\widehat{\mathbf{q}}\mathbf{\cdot}\widehat{\mathbf{p}}+\widehat{\mathbf{p}%
}\mathbf{\cdot}\widehat{\mathbf{q}}\right)  ,\widehat{\mathbf{p}}\right]
\widehat{U}(z)^{-1}=-Y(z) \, , \label{b.4}%
\end{equation}

gives (noting $\widehat{X(}0)=\widehat{\mathbf{q}},$ $\widehat{Y(}%
0)=\widehat{\mathbf{p}}$)
\begin{equation}
\widehat{U}(z)\widehat{\mathbf{q}}\widehat{U}^{-1}(z)=e^{z}\widehat
{\mathbf{q}},\hspace{0.25in}\widehat{U}(z)\widehat{\mathbf{p}}\widehat{U}%
^{-1}(z)=e^{-z}\widehat{\mathbf{p}},\hspace{0.25in}-\infty<z<\infty\, .
\label{b.5}%
\end{equation}
One more definition is needed:
\begin{equation}
\widehat{U}_{\gamma}=\widehat{U}(\ln\left(  \gamma\right)  )=e^{\frac
{i}{2\hbar}\ln\left(  \gamma\right)  \left(  \widehat{\mathbf{q}}%
\mathbf{\cdot}\widehat{\mathbf{p}}+\widehat{\mathbf{p}}\mathbf{\cdot}%
\widehat{\mathbf{q}}\right)  } \, . \label{b.6}%
\end{equation}
This is the unitary operator that generates scale transformations
\begin{equation}
\widehat{U}_{\gamma}A(\widehat{\mathbf{q}}\mathbf{,}\widehat{\mathbf{p}%
})\widehat{U}_{\gamma}^{-1}=A(\gamma\widehat{\mathbf{q}}\mathbf{,}\gamma
^{-1}\widehat{\mathbf{p}}),\hspace{0.25in}0<\gamma<\infty\, . \label{b.7}%
\end{equation}

Now consider the density given by (\ref{3.2}), insert unity $U_{\gamma}%
^{-1}U_{\gamma}=1$, and use cyclic invariance of the trace
\begin{equation}
n(\mathbf{r})=Tr\delta(\widehat{\mathbf{q}}-\mathbf{r})\rho_{e}=TrU_{\gamma
}^{-1}U_{\gamma}\delta(\widehat{\mathbf{q}}-\mathbf{r})\rho=Tr\delta
(\gamma\widehat{\mathbf{q}}-\mathbf{r})\rho_{e\gamma}\,,\nonumber
\end{equation}
or
\begin{equation}
n_{\gamma}\left(  \mathbf{r}\right)  \equiv\gamma^{3}n(\gamma\mathbf{r}%
)=Tr\delta(\widehat{\mathbf{q}}-\mathbf{r})\rho_{e\gamma}\,. \label{b.9}%
\end{equation}
Here $\widehat{\rho}_{e\gamma}$ is the transformed Fermi operator
\begin{equation}
\widehat{\rho}_{e\gamma}\equiv U_{\gamma}\widehat{\rho}_{e}U_{\gamma}%
^{-1}=\left(  e^{\gamma^{-2}\beta\left(  \frac{\widehat{p}^{2}}{2m}-\gamma
^{2}\mu^{(0)}(\gamma\widehat{\mathbf{q}}\mathbf{,}\beta\mid n)\right)
}+1\right)  ^{-1} \label{b.10}%
\end{equation}%
\begin{equation}
\widehat{\rho}_{e\gamma}\equiv U_{\gamma}\widehat{\rho}_{e}(\beta\mid
n)U_{\gamma}^{-1}=\widehat{\rho}_{e}(\gamma^{-2}\beta\mid n_{\gamma})\,.
\label{b.11}%
\end{equation}
Uniqueness of the functional $\mu^{(0)}(\widehat{\mathbf{q}}\mathbf{,}%
\beta\mid n)$ gives the identification
\begin{equation}
\beta\mu(\gamma\widehat{\mathbf{q}}\mathbf{,}\beta\mid n)=\left(  \gamma
^{-2}\beta\right)  \mu(\widehat{\mathbf{q}}\mathbf{,}\gamma^{-2}\beta\mid
n_{\gamma})\,. \label{b.12}%
\end{equation}
Consequently, the statistical operator $\widehat{\rho}_{e}$ obeys the scaling
relation
\begin{equation}
\widehat{\rho}_{e\gamma}(\beta\mid n)=\widehat{\rho}_{e}(\gamma^{-2}\beta\mid
n_{\gamma}). \label{b.13}%
\end{equation}

Confined systems can be represented by including a wall potential in the
external potential. For example, spherically symmetric confinement is
represented $v_{w}=\Theta\left(  R-\widehat{q}\right)  \epsilon_{w}$,
where\ $\epsilon_{w}$ is the potential barrier for escape from the volume $V
=4\pi R^{3}/3$, with the barrier taken arbitrarily large for the hard-wall
limit. Under the scale transformation, the wall potential becomes
\begin{equation}
U_{\gamma}\Theta\left(  R-\widehat{q}\right)  U_{\gamma}^{-1}=\Theta\left(
R-\gamma\widehat{q}\right)  =\Theta\left(  \gamma^{-1}R-\widehat{q}\right)  \,
. \label{b.14}%
\end{equation}
Thus the volume transforms as $V \rightarrow\gamma^{-3} V $.

Consider next the free energy functional (\ref{3.7})
\begin{align}
F^{(0)}\left(  \beta\mid n\right)   &  =\beta^{-1}Tr^{(1)}U_{\gamma}%
^{-1}U_{\gamma}\left[  ..\right] \nonumber\\
&  =\beta^{-1}Tr^{(1)}\left[  \left(  1-\widehat{\rho}_{e\gamma}\right)
\ln\left(  1-\widehat{\rho}_{e\gamma}\right)  +\widehat{\rho}_{e\gamma}%
\ln\widehat{\rho}_{e\gamma}+\beta\gamma^{-2}\frac{\widehat{p}^{2}}{2m}%
\widehat{\rho}_{e\gamma}\right] \nonumber\\
&  =\gamma^{-2}F^{(0)}\left(  \gamma^{-2}\beta\mid n_{\gamma}\right)
\label{b.15}%
\end{align}
The scaling for the energy density and entropy are found in a similar way
\begin{equation}
U^{(0)}(\beta\mid n)=\gamma^{-2}U^{(0)}(\gamma^{-2}\beta\mid n_{\gamma
}),\hspace{0.2in}S^{(0)}(\beta\mid n)=S^{(0)}(\gamma^{-2}\beta\mid n_{\gamma})
\label{b.16}%
\end{equation}

\section{Inequalities from scaling}

\label{apC}

Equations (\ref{4.36}) and (\ref{4.41}) can be combined to give
\begin{equation}
\left(  1-\gamma^{2}\right)  TS^{(0)}(\gamma^{-2}\beta\mid n_{\gamma})\leq
F^{(0)}\left(  \gamma^{-2}\beta\mid n_{\gamma}\right)  -F^{(0)}\left(
\beta\mid n_{\gamma}\right)  \leq\left(  1-\gamma^{2}\right)  TS^{(0)}%
(\beta\mid n_{\gamma}) \, , \label{c.1}%
\end{equation}
and application of the scaling laws gives
\begin{equation}
\frac{\left(  1-\gamma^{2}\right)  }{\gamma^{2}}TS^{(0)}(\beta\mid
n)\leq\left(  F^{(0)}\left(  \beta\mid n\right)  -F^{(0)}\left(  \gamma
^{2}\beta\mid n\right)  \right)  \leq\frac{\left(  1-\gamma^{2}\right)
}{\gamma^{2}}TS^{(0)}(\gamma^{2}\beta\mid n) \, . \label{c.2}%
\end{equation}
This implies 
\begin{equation}
S^{(0)}(\beta\mid n)\lesseqgtr S^{(0)}(\gamma^{2}\beta\mid n),\hspace
{0.2in}\gamma\lesseqgtr1 \label{c.3}%
\end{equation}
or, equivalently,
\begin{equation}
S^{(0)}(\beta\mid n)\lesseqgtr S^{(0)}(\beta\mid n_{\lambda}),\hspace
{0.2in}\gamma\lesseqgtr1 \, . \label{c.4}%
\end{equation}

Next, writing (\ref{4.36}) in terms of the energy and density
\begin{equation}
U^{(0)}\left(  \beta\mid n_{\gamma}\right)  -U^{(0)}\left(  \gamma^{-2}%
\beta\mid n_{\gamma}\right)  \leq T\left(  S^{(0)}(\beta\mid n_{\gamma
})-S^{(0)}(\gamma^{-2}\beta\mid n_{\gamma})\right)  \,, \label{c.5}%
\end{equation}
or with scaling
\begin{equation}
U^{(0)}\left(  \beta\mid n_{\gamma}\right)  -\gamma^{2}U^{(0)}\left(
\beta\mid n\right)  \leq T\left(  S^{(0)}(\beta\mid n_{\gamma})-S^{(0)}%
(\beta\mid n)\right)  \, . \label{c.6}%
\end{equation}
Then (\ref{c.3}) and (\ref{c.4}) give
\begin{equation}
U^{(0)}\left(  \beta\mid n_{\gamma}\right)  \lesseqgtr\gamma^{2}U^{(0)}\left(
\beta\mid n\right)  ,\hspace{0.2in}\gamma\gtreqless1 \label{c.7}%
\end{equation}
Expressing these results in terms of the same density by scaling gives the
equivalent forms
\begin{equation}
U^{(0)}\left(  \gamma^{2}\beta\mid n\right)  \lesseqgtr U^{(0)}\left(
\beta\mid n\right)  ,\hspace{0.2in}\gamma\gtreqless1. \label{c.8}%
\end{equation}

\end{document}